\documentclass[10pt]{article}
\usepackage{opex2}
\usepackage{epsfig}
\usepackage{amsmath}
\usepackage{amssymb}

\begin{document}

\title{Multi-mode photonic crystal fibers for VCSEL based data transmission}

\author{N.~A. Mortensen,$^{1*}$ M. Stach,$^2$ J. Broeng,$^1$ A. Petersson,$^1$ H.~R. Simonsen,$^1$ and R. Michalzik$^2$}

\address{$^1$Crystal Fibre A/S, Blokken 84, DK-3460 Birker\o d, Denmark\\
$^2$University of Ulm, Optoelectronics Department,\\ Albert-Einstein-Allee 45, D-89069 Ulm, Germany }

\email{$^*$nam@crystal-fibre.com}

\begin{abstract}
Quasi error-free $10\,{\rm Gbit/s}$ data transmission is
demonstrated over a novel type of $50\,{\rm \mu m}$ core diameter
photonic crystal fiber with as much as $100\,{\rm m}$ length.
Combined with $850$\,nm VCSEL sources, this fiber is an
attractive alternative to graded-index multi-mode fibers for
datacom applications. A comparison to numerical simulations suggests that the high bit-rate may be partly explained by inter-modal diffusion.
\end{abstract}


%\bibliographystyle{ol_titles}
%\bibliography{/mnt/nam/papers/PCF}

\section{Introduction}
Optical datacom as employed for the high-speed interconnection of
electronic sub-systems has rapidly gained importance over the
past years. Vertical-cavity surface-emitting lasers (VCSELs)
emitting in the $850$\,nm wavelength regime and simple
step-index fibers or graded-index fibers are preferred key
components for low-cost link solutions \cite{michalzik2001}.
Whereas, due to strong inter-modal dispersion, the use of the
former fiber type is limited to link lengths of some meters at
Gbit/s data rates, fabrication of the latter requires supreme
control over the refractive index profile, especially in
optimized $50\,{\rm \mu m}$ core diameter fibers enabling up to
$300\,{\rm m}$ serial transmission of $10\,{\rm Gbit/s}$ signals.
Since optical interconnect requirements move toward higher speed
over shorter distances, the availability of an easily
manufacturable, yet high-speed capable fiber medium would be very
beneficial. In this paper, we report on the properties of a new
type of multi-mode photonic crystal fiber (PCF) with relatively
simple waveguide geometry and demonstrate $850$\,nm data
transmission at $10\,{\rm Gbit/s}$ over a length of
$L=100\,{\rm m}$. For a recent review of photonic crystal fibers
we refer to Ref.~\cite{russell2003} and references therein.

\section{Fiber design}\label{design}

\begin{figure}[t!]
\begin{center}

\epsfig{file=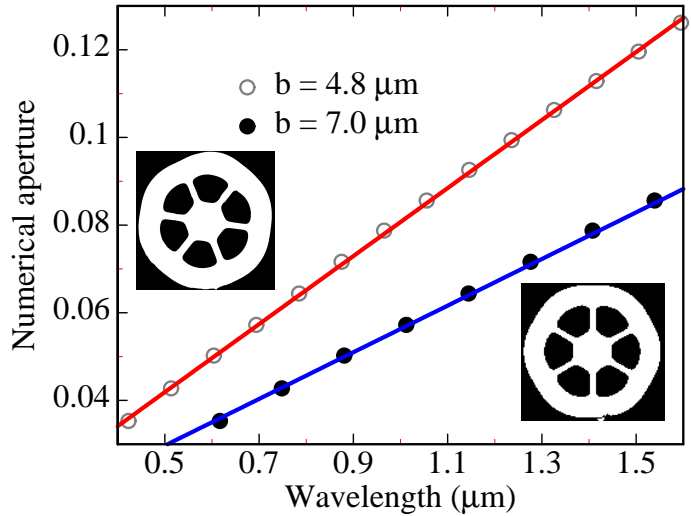, width=0.7\textwidth,clip}
\end{center}
\caption{Simulated $\rm NA$ for the $33\,{\rm \mu m}$ core PCF
(upper left inset) with bridges of width $b\simeq 4.8\,{\rm \mu m}$
and the $50\,{\rm \mu m}$ core PCF (lower right inset) with
bridges of width $b\simeq 7.0\,{\rm \mu m}$. Note the different
scale for the two insets. } \label{fig1}
\end{figure}

The design of the new multi-mode photonic crystal fiber is
illustrated in the insets of Fig.~\ref{fig1} which show optical
micrographs of the fiber cross-sections. The fibers are made from
a single material (light regions), and they comprise a solid,
pure silica core suspended in air (dark regions) by narrow silica
bridges of width $b$.

There is a large degree of freedom in engineering the optical properties and still get fiber designs of practical interest from a fabrication point of view. The properties may be tailored by adjusting parameters such as the size and shape of
the core, the dimensions and number of silica bridges, or the
fiber material. The numerical aperture (NA) of this type of PCF
is essentially determined by the width of the silica bridges
relative to the wavelength $\lambda$ as numerically demonstrated in Fig.~\ref{fig1}. Here, we focus on two fibers
with $33\,{\rm \mu m}$ and $50\,{\rm \mu m}$ core diameter and
bridge widths of $b=4.8 \,{\rm \mu m}$ and $7.0 \,{\rm \mu m}$,
respectively, yielding NAs of around $0.07$ and $0.05$ at a wavelength of 850\,nm.

Despite the zero-index step between the core and the bridges, the
fiber is capable of guiding light with good confinement to the
multi-mode core. This is illustrated by the near-field intensity distributions for both the  $33\,{\rm \mu m}$
core PCF (Fig.~\ref{fig6}) as well as the
$50\,{\rm \mu m}$ core PCF (the inset in Fig.~\ref{fig7}).

We find that the fibers can be cleaved and spliced with commercially available equipment and typically, the fibers have an attenuation of the order $50$ dB/km at 850 nm for typical bending radii such as 16 cm.

\section{Transmission experiments}\label{transmission}

Assuming worst-case conditions \cite{agrawal_c}, we estimate from
the above NA-values a bit rate-length product of around
$350\, {\rm MBit/s \times  km}$ for the $50\,{\rm \mu m}$ fiber,
whereas the $33\,{\rm\mu m}$ sample should have around
$180\, {\rm MBit/s \times  km}$. In what follows we examine the
transmission properties of such PCFs with a length of
$L=100\,{\rm m}$.

\subsection{Small-signal transfer function and DMD}

In order to get a first indication of the fibers' expected
transmission bandwidths, we have determined the small-signal
frequency responses with a scalar network analyzer. As optical
source, standard 850\,nm GaAs based VCSELs have been employed.
The $12\,{\rm \mu m}$ active diameter, oxide-confined devices
show transverse multi-mode emission with a root mean square
spectral width of less than $0.4\,{\rm nm}$ even under
modulation. The lasing threshold current amounts to
$1.8\,{\rm mA}$ and the bias current for the small-signal as
well as data transmission experiments was chosen as $9\,{\rm
 mA}$, where the 3-dB bandwidth is $8.6\,{\rm  GHz}$. At the
receiving end, a multi-mode fiber pigtailed InGaAs
pin-photo-receiver with above $8\,{\rm GHz}$ bandwidth was used.

\begin{figure}[t!]
\begin{center}
\epsfig{file=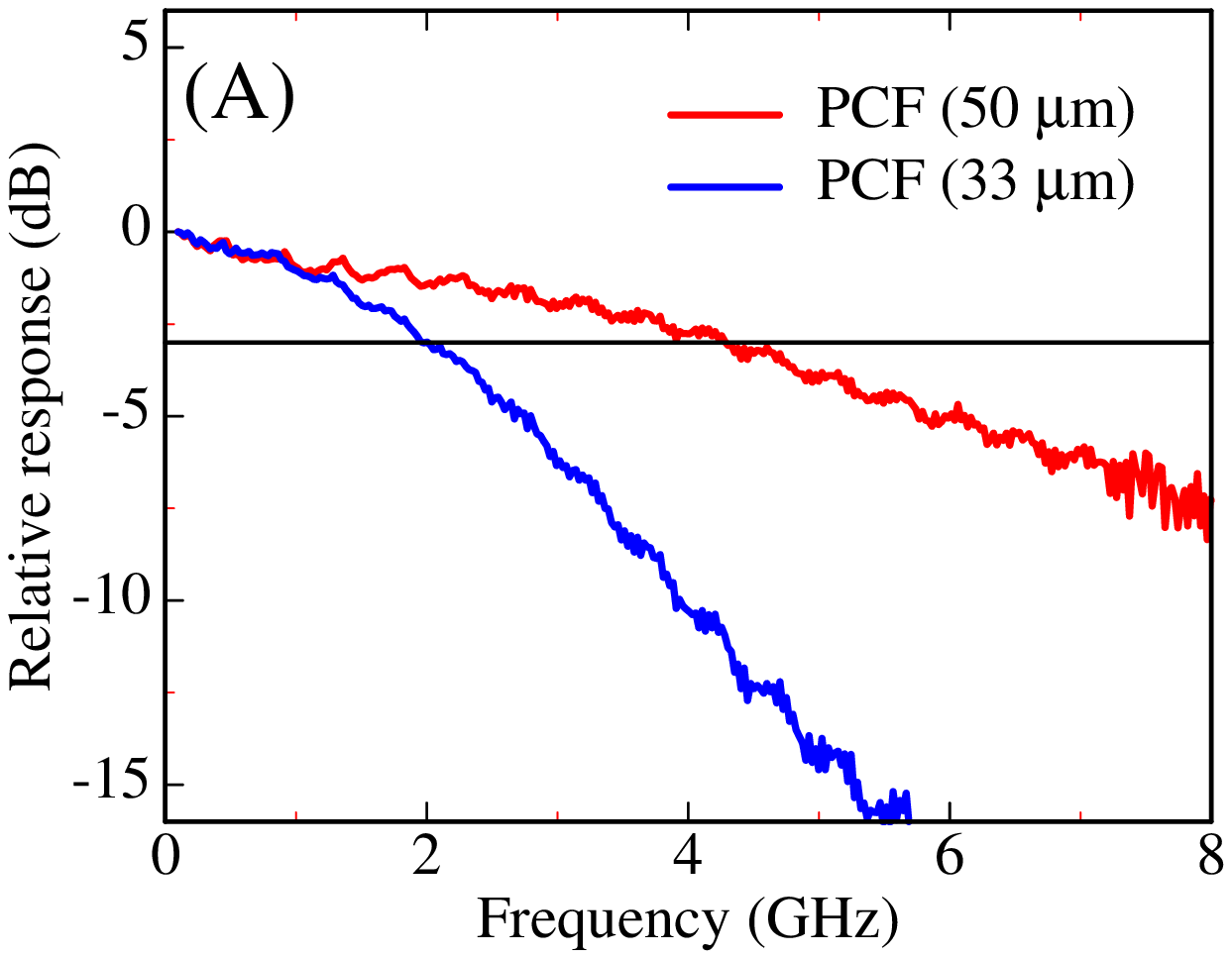, height=0.36\textwidth,clip}\epsfig{file=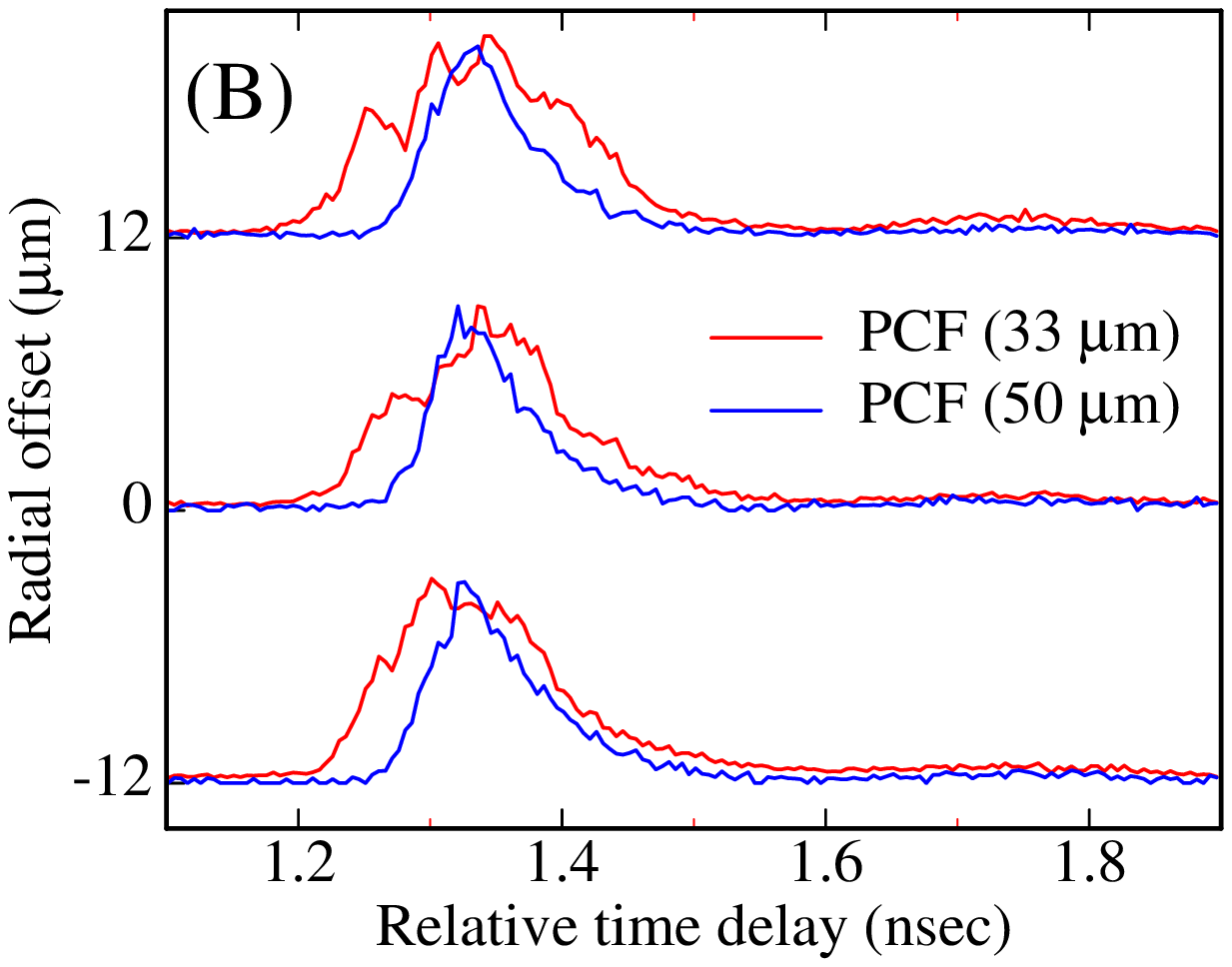, height=0.36\textwidth,clip}
\end{center}
\caption{Panel (A) shows small-signal frequency responses at
$850\,{\rm nm}$ for a length $L=100\,{\rm m}$ for the two PCFs
illustrated in Fig.~\ref{fig1}. Panel (B) shows normalized DMD
plots for both fibers at offset positions of $-12$, $0$, and
$12\,{\rm\mu m}$. } \label{fig2}
\end{figure}

Panel (A) of Fig.~\ref{fig2} depicts the relative responses of
both PCF samples. The $33$ and $50\,{\rm \mu m}$ core PCFs show a bit rate-length product of $B_T\times L\sim 500\,{\rm
Mbit/s \times km}$ and $\sim 1000\,{\rm
Mbit/s \times km}$, respectively. These figures are significantly larger than expected from the corresponding NAs. In the
next section we extend the NA estimations and show simulations of
the modal time delays for the two PCFs.

\begin{figure}[b!]
\begin{center}
\epsfig{file=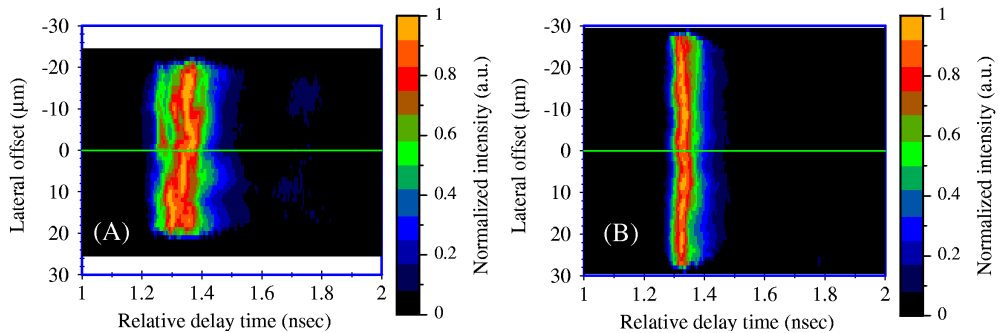, width=0.99\textwidth,clip}
\end{center}
\caption{Normalized DMD plots at variable offset positions. Panels (A) and (B) show results for the $33\,{\rm \mu m}$ and the $50\,{\rm\mu m}$ PCFs, respectively.} \label{fig3}
\end{figure}

In order to get quantitative insight into the modal delay
properties, we have determined the PCFs differential mode delay
(DMD) characteristics, see Panel (B) of Fig.~\ref{fig2}. Here, a $5\,\mu$m core diameter single-mode fiber is scanned over the PCF input at a distance of about $10\,\mu$m in accordance with the IEC pre-standard 60793-1-49, Sect.\ 3.3. The impulse response at the output end is recorded for each offset position using an optical sampling oscilloscope with a fiber input compatible to $62.5\,\mu$m core diameter multi-mode fibers. A gain-switched 850\,nm single-mode VCSEL delivering pulses with
less than 40\,ps full width at half maximum is employed for this purpose \cite{RaM-APOC}. 
Panel (B) illustrates some of the results. It is seen
that the output pulses of the 50\,$\mu$m fiber are rather
narrow and virtually independent of the offset position. On the
other hand, those of the $33\,{\rm\mu m}$ sample show larger
variability and are up to twice as broad, which well supports the
above observations. Figure~\ref{fig3} shows two-dimensional color-coded representations of the full data.

\begin{figure}[t!]
\begin{center}
\epsfig{file=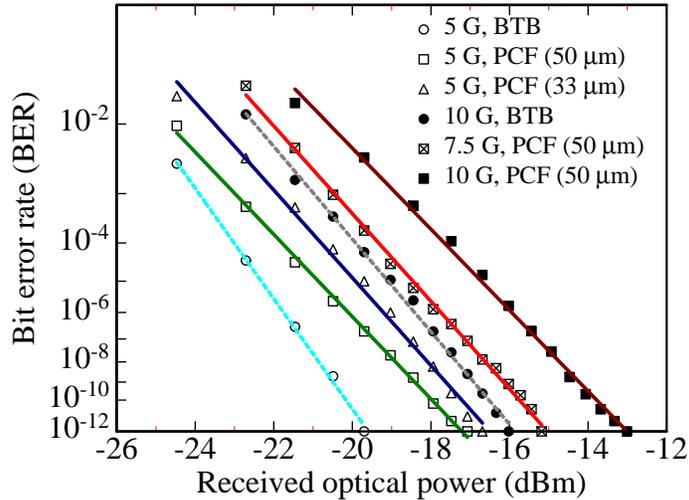, width=0.7\textwidth,clip}
\end{center}
\caption{ BER characteristics for both 100\,m-long PCFs at data
rates of $5$, $7.5$, and $10\,{\rm Gbit/s}$. } \label{fig4}
\end{figure}

\subsection{Digital data transmission}
Data transmission experiments have been carried out under
non-return-to-zero $2^7 - 1$ word length pseudo-random bit sequence
modulation using the aforementioned multi-mode VCSEL
driven with $0.9\,{\rm  V}$
peak-to-peak voltage. Figure \ref{fig4} summarizes obtained bit
error rate (BER) curves. With the smaller core diameter fiber, up
to $5\,{\rm Gbit/s}$ could be transmitted without indication of
a BER floor. The power penalty versus back-to-back (BTB)
operation is about 3\,dB at a BER of $10^{-12}$. On the other
hand, the $50\,{\rm \mu m}$ fiber even enables
$10\,{\rm Gbit/s}$ transmission over $L=100\,{\rm m}$ length
with only $2.9\,{\rm dB}$ power penalty. The observed increase
in data rate is in full agreement with the small-signal and DMD
measurement results.

\begin{figure}[h!]
\begin{center}
\epsfig{file=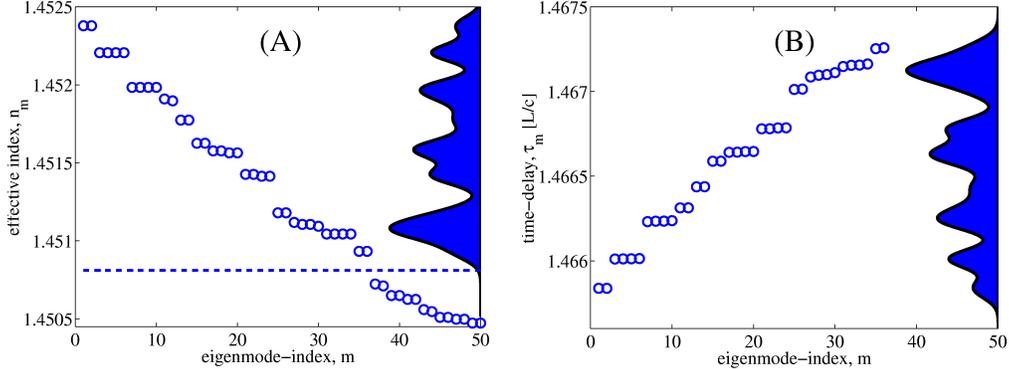, width=1\textwidth,clip}
\end{center}
\caption{ Panel (A) shows the effective indices of the $M=36$
guided eigenmodes at $\lambda=850\,{\rm nm}$ in the $33\,{\rm \mu m}$ core PCF (see upper left inset of Fig.~\ref{fig1}).
The horizontal dashed line indicates the cladding index $n_{\rm
cl}$ corresponding to the experimentally measured $\rm NA$. The
filled curve shows the distribution $P(n_m)$ (the projection of
the data onto the $y$-axis). Panel (B) shows the corresponding
time-delays $\tau_m$ and the distribution $P(\tau_m)$.  }
\label{fig5}
\end{figure}

\section{Simulations}

We use a plane-wave method~\cite{johnson2001} to calculate the propagation
constant $\beta_m = n_m\omega /c$ of the $m$th eigenmode where $n_m$ is the effective index, $\omega$ the angular frequency, and $c$ the vacuum velocity of light. For the refractive index profile we use optical micrographs
transformed to one-bit format representing
the two-component composite air-silica structure and for the
refractive index we use a Sellmeier expression for $n(\omega)$
in silica and $n=1$ in air. The simulation of Maxwell's equations
for a given $\omega$ provides us with sets of propagation
constants $\{\beta_m\}$ and eigenfields $\{{\boldsymbol E}_m\}$
where $m=1,2,3,\ldots M$ with $M$ as the number of guided eigenmodes.
We determine $M$ from the experimentally measured ${\rm NA}$
which we transform to an effective cladding index
$n_{\rm cl}$. The number of guided eigenmodes $M$ then follows
from the requirement that $n_M > n_{\rm cl} \geq n_{M+1}$.

The delay-times (or group-delays) are given by $\tau_m = L
\partial \beta_m/\partial\omega$ (we calculate the group velocity
by the approach described in Ref.~\cite{laegsgaard2003}).
The variation with $m$ usually sets the limit on the bit rate and in that case the bit rate-length
product is given by~\cite{agrawal_c,ghatak1998}

\begin{equation}
B_T\times L\simeq L/\Delta T\;,\;\Delta T \approx 2 \sqrt{\big< \{ \delta^2\tau_m\}\big>},\;\;\delta\tau_m =\tau_m-\big<\{\tau_m\}\big>,
\end{equation}
Here, we use the second moment calculated from the full statistics to characterize the width $\Delta T$ of the distribution $P(\tau_m)$. For the estimate of the bit-rate the eigenmodes are thus weighted equally corresponding to an assumption of uniform launch and attenuation. In literature one often finds the estimate $\Delta T\approx \max\{\tau_m\}-\min\{\tau_m\}$ \cite{agrawal_c,ghatak1998} and in the ray-optical picture
$\max\{\tau_m\}$ can be expressed in terms of the
$\rm NA$ in analogy to our estimations in section~\ref{transmission} based on the $\rm NA$. However, for a sufficiently low number of guided modes the beginning break-down of geometrical optics calls for estimates based on the full statistics.

Figure~\ref{fig5} shows results at $\lambda=850\,{\rm nm}$ for the $33\,{\rm \mu m}$ core PCF (see upper left inset in
Fig.~\ref{fig1}). Experimentally, this fiber is found to have an
$\rm NA\simeq 0.07$ and the corresponding effective cladding
index is indicated by the dashed line in panel (A). For the given
core size this results in $M=36$ eigenmodes that are guided.
Panel (B) shows the results for the time-delays with the filled
curve showing the distribution $P(\tau_m)$ (the projection of the data onto the $y$-axis) calculated from a superposition of
Gaussians with a width given by the mean level spacing
$(\tau_M-\tau_1)/(M-1)$. We have $\Delta T \simeq 0.00087 \times L/c$
corresponding to $B_T\times L\simeq  344\, {\rm MBit/s}\times{\rm  km}$ which as expected is somewhat larger than the NA-estimate. The experimentally observed value is approximately $50\%$ larger. It is wellknown that both non-uniform loss and attenuation as well as inter-modal diffusion tends to narrow the spread in time-delays. The DMD plots in Fig.~\ref{fig3} supports the presence of inter-modal diffusion and its dominance over both the excitation conditions as well as variations in modal attenuation. It is thus likely that the enhanced bit-rate length product originates from inter-modal diffusion. One could speculate that stress could modify the index-profile in the silica core and that this in turn could modify the time-delay distribution similarly to the situation in graded-index profiles. However, as we shall see such a hypothesis is not supported by near-field studies.

\begin{figure}[t!]
\begin{center}
\epsfig{file=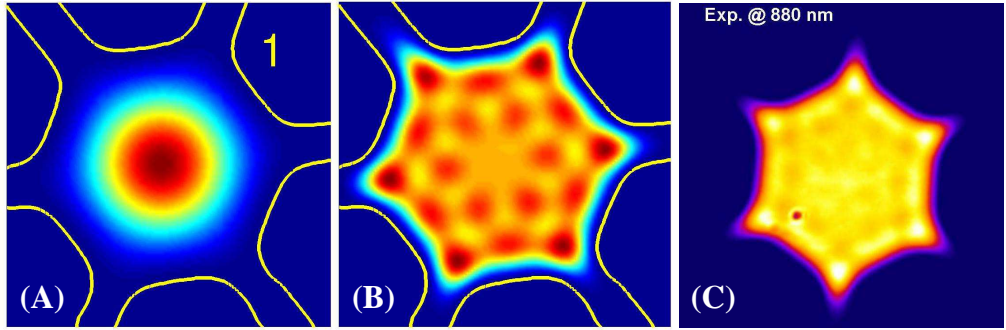, width=1\textwidth,clip}
\end{center}
\caption{Intensity distributions at $\lambda=850\,{\rm nm}$ in the $33\,{\rm \mu m}$ PCF (see upper left inset in Fig.~\ref{fig1}). Panel (A) shows the first ($m=1$) eigenmode (see http://asger.hjem.wanadoo.dk/mm.gif to view the other $M=36$ guided eigenmodes, 700 Kbyte). Panel (B) shows the average eigenfield intensity which agrees well with the experimentally observed near-field intensity shown in Panel (C). In Panels (A) and (B) the contour lines indicate the air-silica interfaces.}
\label{fig6}
\end{figure}

The electric field ${\boldsymbol E}$ is constructed by a linear
combination of the eigenfields. For a not too narrow linewidth
of the light source we may neglect cross-terms in $| {\boldsymbol E}|^2$
and for uniform launch and attenuation we thus expect to measure an intensity
distribution proportional to the average eigenfield intensity,
i.e., $|{\boldsymbol E}|^2 \approx M^{-1}\sum_m^M |{\boldsymbol E}_m|^2 $. The same will be the case for arbitrary launch and strong inter-modal diffusion. Figure~\ref{fig6} shows the
eigenfield intensities with spatial patterns characteristic for
a close-to-hexagonal symmetry. The average eigenfield intensity in Panel (B) compares well to the experimentally measured
near-field intensity in Panel (C). Together with the DMD measurements this correspondence agrees well with a picture of inter-modal diffusion which tends to populate the modes uniformly.

\begin{figure}[t!]
\begin{center}
\epsfig{file=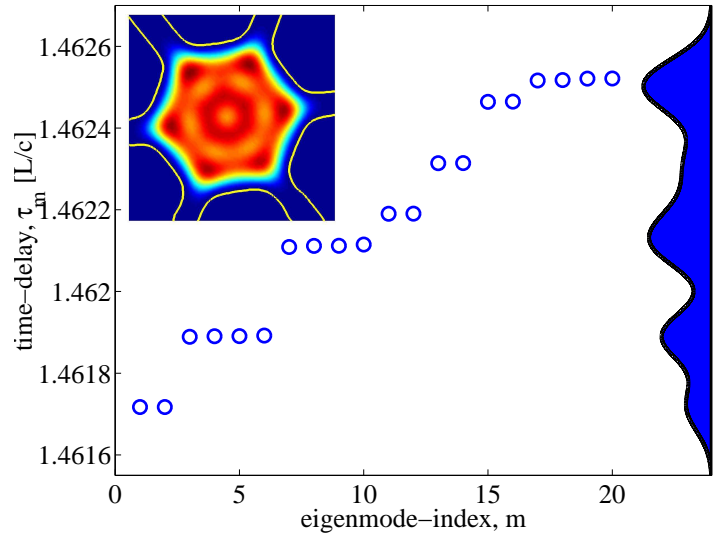, width=0.7\textwidth,clip}
\end{center}
\caption{Time-delays of the $M=20$ guided eigenmodes in the
$50\,{\rm \mu m}$ PCF (see lower right inset in Fig.~\ref{fig1}).
The filled curve shows the distribution $P(\tau_m)$ and the inset shows the simulated average eigenfield intensity with contour lines indicating the air-silica interfaces.} \label{fig7}
\end{figure}

The eigenmodes fall into different groups with different
degeneracies (these degeneracies are slightly lifted due to a
weakly broken symmetry in the real fiber) as evident from both
the effective index in panel (A) of Fig.~\ref{fig5} as well as
the intensity plots (click panel (A) in Fig.~\ref{fig6}). The first two eigenmodes
($m=1,2$) are the doubly degenerate fundamental mode
corresponding to the two polarization states of the fundamental
mode in standard fibers and from a practical point of view they
can be considered polarization states though the
``$x$-polarization'' in principle has a very small $y$-component and vice versa.

For the $50\,{\rm \mu m}$ PCF (see lower right inset of
Fig.~\ref{fig1}) with ${\rm NA}\simeq 0.05$ we have carried out
the same analysis of the effective index and found that $M=20$
eigenmodes are guided. Since $M$ increases with both increasing
$\rm NA$ and core size, $M$ can be low even for a large core as long as the $\rm NA$ is not too high. Figure~\ref{fig7} shows
results for the time-delay which as expected has a more narrow
distribution compared to the results for the PCF with the
$33\,{\rm \mu m}$ core, see panel (B) of Fig.~\ref{fig5}. The
width $\Delta T \simeq 0.00054 \times L/c$ corresponds to
$B_T\times L\simeq  559\, {\rm MBit/s}\times{\rm  km}$. The experimental value is larger by more than $70\%$ which is attributed to inter-modal diffusion.

\section{Conclusions}
For the first time, quasi error-free transmission of 10\,Gbit/s
digital data signals over a multi-mode photonic crystal fiber
with 50\,$\mu$m core diameter and as much as 100\,m length
has been demonstrated. With some optimizations concerning design
and fabrication, these PCFs show good prospects as an alternative
to graded-index fibers in optical datacom environments. Comparing to numerical simulations indicates that the high bit-rate may be partly supported by inter-modal diffusion.


\begin{thebibliography}{1}
\newcommand{\enquote}[1]{`#1'}

\bibitem{michalzik2001}
R.~Michalzik, K.~J. Ebeling, M.~Kicherer, F.~Mederer, R.~King, H.~Unold, and
  R.~Jager, ``High-performance VCSELs for optical data links,'' IEICE T. Electron. {\bf E84C}, 629 (2001).

\bibitem{russell2003}
P.~Russell, ``Review: Photonic Crystal Fibers,'' Science {\bf 299}, 358 (2003).

\bibitem{agrawal_c}
G.~P. Agrawal, {\it Fiber-Optic Communication Systems} (Wiley \& Sons, New York, 1997).

\bibitem{RaM-APOC}
R.~Michalzik, F.~Mederer, H.~Roscher, M.~Stach, H.~Unold, D.~Wiedenmann,
  R.~King, M.~Grabherr, and E.~Kube, ``Design and communication applications of short-wavelength VCSELs,'' Proc. SPIE {\bf 4905}, 310 (2002).

\bibitem{johnson2001}
S.~G. Johnson and J.~D. Joannopoulos, ``Block-iterative frequency-domain methods for \uppercase{M}axwell's equations in a planewave basis,'' Opt. Express {\bf 8}, 173 (2001),\\ http://www.opticsexpress.org/abstract.cfm?URI=OPEX-8-3-173 

\bibitem{laegsgaard2003}
J.~L{\ae}gsgaard, A.~Bjarklev, and S.~E.~B. Libori, ``Chromatic dispersion in photonic crystal fibers: fast and accurate scheme for calculation,'' J. Opt. Soc. Am. B {\bf 20}, 443 (2003).

\bibitem{ghatak1998}
A.~K. Ghatak and K.~Thyagarajan, {\it Introduction to Fiber Optics} (Cambridge University Press, Cambridge, 1998).

\end{thebibliography}
\end{document}